\documentclass[manuscript]{aastex}

\slugcomment{Not to appear in Nonlearned J., 45.}

\shorttitle{OBSERVATION OF TeV GAMMA RAYS FROM THE FERMI SOURCES}
\shortauthors{Amenomori et al.}

\begin{document}


\title{OBSERVATION OF TeV GAMMA RAYS \\
FROM THE FERMI BRIGHT GALACTIC SOURCES \\
WITH THE TIBET AIR SHOWER ARRAY}


\author{ 
M.~Amenomori\altaffilmark{1}, X.~J.~Bi\altaffilmark{2},
D.~Chen\altaffilmark{3}, S.~W.~Cui\altaffilmark{4},
Danzengluobu\altaffilmark{5}, L.~K.~Ding\altaffilmark{2},
X.~H.~Ding\altaffilmark{5}, C.~Fan\altaffilmark{6,2},
C.~F.~Feng\altaffilmark{6}, Zhaoyang Feng\altaffilmark{2},
Z.~Y.~Feng\altaffilmark{7}, X.~Y.~Gao\altaffilmark{8},
Q.~X.~Geng\altaffilmark{8}, Q.~B.~Gou\altaffilmark{2},
H.~W.~Guo\altaffilmark{5}, H.~H.~He\altaffilmark{2},
M.~He\altaffilmark{6}, K.~Hibino\altaffilmark{9},
N.~Hotta\altaffilmark{10}, Haibing~Hu\altaffilmark{5},
H.~B.~Hu\altaffilmark{2}, J.~Huang\altaffilmark{2},
Q.~Huang\altaffilmark{7}, H.~Y.~Jia\altaffilmark{7},
L.~Jiang\altaffilmark{8,2}, F.~Kajino\altaffilmark{11},
K.~Kasahara\altaffilmark{12}, Y.~Katayose\altaffilmark{13},
C.~Kato\altaffilmark{14}, K.~Kawata\altaffilmark{3},
Labaciren\altaffilmark{5}, G.~M.~Le\altaffilmark{15},
A.~F.~Li\altaffilmark{6}, H.~C.~Li\altaffilmark{4,2},
J.~Y.~Li\altaffilmark{6}, C.~Liu\altaffilmark{2},
Y.-Q.~Lou\altaffilmark{16}, H.~Lu\altaffilmark{2},
X.~R.~Meng\altaffilmark{5}, K.~Mizutani\altaffilmark{12,17},
J.~Mu\altaffilmark{8}, K.~Munakata\altaffilmark{14},
H.~Nanjo\altaffilmark{1},
M.~Nishizawa\altaffilmark{18}, M.~Ohnishi\altaffilmark{3},
I.~Ohta\altaffilmark{19}, S.~Ozawa\altaffilmark{12},
T.~Saito\altaffilmark{20}, T.~Y.~Saito\altaffilmark{21},
M.~Sakata\altaffilmark{11}, T.~K.~Sako\altaffilmark{3},
M.~Shibata\altaffilmark{13}, A.~Shiomi\altaffilmark{22},
T.~Shirai\altaffilmark{9}, H.~Sugimoto\altaffilmark{23},
M.~Takita\altaffilmark{3}, Y.~H.~Tan\altaffilmark{2},
N.~Tateyama\altaffilmark{9}, S.~Torii\altaffilmark{12},
H.~Tsuchiya\altaffilmark{24}, S.~Udo\altaffilmark{9},
B.~Wang\altaffilmark{2}, H.~Wang\altaffilmark{2},
Y.~Wang\altaffilmark{2}, Y.~G.~Wang\altaffilmark{6},
H.~R.~Wu\altaffilmark{2}, L.~Xue\altaffilmark{6},
Y.~Yamamoto\altaffilmark{11}, C.~T.~Yan\altaffilmark{25},
X.~C.~Yang\altaffilmark{8}, S.~Yasue\altaffilmark{26},
Z.~H.~Ye\altaffilmark{27}, G.~C.~Yu\altaffilmark{7},
A.~F.~Yuan\altaffilmark{5}, T.~Yuda\altaffilmark{9},
H.~M.~Zhang\altaffilmark{2}, J.~L.~Zhang\altaffilmark{2},
N.~J.~Zhang\altaffilmark{6}, X.~Y.~Zhang\altaffilmark{6},
Y.~Zhang\altaffilmark{2}, Yi~Zhang\altaffilmark{2},
Ying~Zhang\altaffilmark{7,2}, Zhaxisangzhu\altaffilmark{5} and
X.~X.~Zhou\altaffilmark{7}\\ (The Tibet AS$\gamma$ Collaboration) }


\altaffiltext{1}{Department of Physics, Hirosaki University, Hirosaki 036-8561, Japan.}
\altaffiltext{2}{Key Laboratory of Particle Astrophysics, Institute of High Energy Physics, Chinese Academy of Sciences, Beijing 100049, China.}
\altaffiltext{3}{Institute for Cosmic Ray Research, University of Tokyo, Kashiwa 277-8582, Japan.}
\altaffiltext{4}{Department of Physics, Hebei Normal University, Shijiazhuang 050016, China.}
\altaffiltext{5}{Department of Mathematics and Physics, Tibet University, Lhasa 850000, China.}
\altaffiltext{6}{Department of Physics, Shandong University, Jinan 250100, China.}
\altaffiltext{7}{Institute of Modern Physics, SouthWest Jiaotong University, Chengdu 610031, China.}
\altaffiltext{8}{Department of Physics, Yunnan University, Kunming 650091, China.}
\altaffiltext{9}{Faculty of Engineering, Kanagawa University, Yokohama 221-8686, Japan.}
\altaffiltext{10}{Faculty of Education, Utsunomiya University, Utsunomiya 321-8505, Japan.}
\altaffiltext{11}{Department of Physics, Konan University, Kobe 658-8501, Japan.}
\altaffiltext{12}{Research Institute for Science and Engineering, Waseda University, Tokyo 169-8555, Japan.}
\altaffiltext{13}{Faculty of Engineering, Yokohama National University, Yokohama 240-8501, Japan.}
\altaffiltext{14}{Department of Physics, Shinshu University, Matsumoto 390-8621, Japan.}
\altaffiltext{15}{National Center for Space Weather, China Meteorological Administration, Beijing 100081, China.}
\altaffiltext{16}{Physics Department and Tsinghua Center for Astrophysics, Tsinghua University, Beijing 100084, China.}
\altaffiltext{17}{Saitama University, Saitama 338-8570, Japan.}
\altaffiltext{18}{National Institute of Informatics, Tokyo 101-8430, Japan.}
\altaffiltext{19}{Sakushin Gakuin University, Utsunomiya 321-3295, Japan.}
\altaffiltext{20}{Tokyo Metropolitan College of Industrial Technology, Tokyo 116-8523, Japan.}
\altaffiltext{21}{Max-Planck-Institut f\"ur Physik, M\"unchen D-80805, Deutschland.}
\altaffiltext{22}{College of Industrial Technology, Nihon University, Narashino 275-8576, Japan.}
\altaffiltext{23}{Shonan Institute of Technology, Fujisawa 251-8511, Japan.}
\altaffiltext{24}{RIKEN, Wako 351-0198, Japan.}
\altaffiltext{25}{Institute of Disaster Prevention Science and Technology, Yanjiao 065201, China.}
\altaffiltext{26}{School of General Education, Shinshu University, Matsumoto 390-8621, Japan.}
\altaffiltext{27}{Center of Space Science and Application Research, Chinese Academy of Sciences, Beijing 100080, China.}


\begin{abstract}
Using the Tibet-III air shower array, we search for TeV $\gamma$-rays
from 27 potential Galactic sources in the early list of bright sources
obtained by the {\it Fermi} Large Area Telescope at energies above
100~MeV.  Among them, we observe 7 sources instead of the expected 0.61
sources at a significance of 2$\sigma$ or more excess.  The chance
probability from Poisson statistics would be estimated to be
3.8$\times$10$^{-6}$.
If the excess distribution observed by the Tibet-III array has a
density gradient toward the Galactic plane, the expected number of
sources may be enhanced in chance association.  Then, the
chance probability rises slightly, to 1.2$\times$10$^{-5}$, based on a
simple Monte Carlo simulation.
These low chance probabilities clearly show that
the {\it Fermi} bright Galactic sources have statistically
significant correlations with TeV $\gamma$-ray excesses.  We also find
that all 7 sources are associated with pulsars, and 6 of them are
coincident with sources detected by the Milagro experiment at a
significance of 3$\sigma$ or more at the representative energy of 35~TeV.
The significance maps observed by the Tibet-III air shower array
around the {\it Fermi} sources, which are coincident with the Milagro
$\ge$3$\sigma$ sources, are consistent with the Milagro observations.
This is the first result of the northern sky survey of the {\it Fermi}
bright Galactic sources in the TeV region.
\end{abstract}





\section{Introduction}
The {\it Fermi Gamma-ray Space Telescope} ({\it Fermi}), succeeding
the Energetic Gamma Ray Experiment (EGRET), was launched in June 2008
to cover the energy range of 20~MeV to 300~GeV, with a sensitivity
approximately a hundred times better than that of the EGRET.  The
Large Area Telescope (LAT) on board the {\it Fermi} surveyed
the entire sky for 3 months, after which the 205 most significant sources
were published in a bright source list above 100~MeV at a significance
greater than $\sim$10$\sigma$ \cite{Abd09a}.  Remarkably, this survey detected many new
$\gamma$-ray pulsars.  A typical 95\%
uncertainty radius of source position in this list is approximately
10$\arcmin$ and the maximum is 20$\arcmin$; these values are greatly improved
compared to those of the EGRET.  This provides a more accurate,
unbiased search for common sources across multi wavelengths,
compared with the EGRET era.  Recently, the Milagro experiment
observed 14 of the 34 {\it Fermi} sources selected from
the list at a false-positive significance of 3$\sigma$ or more at the
representative energy of 35~TeV \cite{Abd09b}.

In this paper, we report on a search for TeV $\gamma$-ray sources in
the {\it Fermi} bright source list with the Tibet-III air shower array
(Tibet-III array).  We also discuss simple statistical tests for our
results and possible coincidences with the Milagro observations.

\section{Tibet-III Air Shower Array} \label{s-2}
The Tibet air shower array has been operating at Yangbajing Cosmic Ray
Observatory ($90\fdg522$ east, $30\fdg102$ north; 4300~m above sea
level) in Tibet, China, since 1990 \cite{Ame92}.  We observe cosmic
rays and $\gamma$-rays using the extensive air shower technique of a
scintillation detector array with a duty cycle of about 24 hours every
day regardless of weather conditions, and with a wide field of view of
about 2~sr.  These capabilities give us an unbiased survey of the
northern sky.  After several upgrades, the Tibet-III array 
was completed and started collecting data in late 1999.  This
array consists of 533 plastic scintillation detectors of 0.5~m$^2$
placed at grid point 7.5~m apart, and its coverage area
is approximately 22,050 m$^{2}$ \cite{Ame03}.  Each detector, 
called a fast-timing (FT) detector, has an FT photomultiplier tube to
collect scintillation photons. The number of air shower particles and
the arrival timing of particles at each detector are recorded, allowing us to
estimate primary cosmic ray or $\gamma$-ray direction and
energy for each air shower.  The systematic uncertainty of the absolute
energy scale observed by the Tibet-III array in the multi-TeV region is
calibrated to be less than $\pm$12\% using the Moon's shadow
observation \cite{Ame09}.  The single event angular resolution is
estimated to be $0\fdg9$ for modal energy 3~TeV, although it depends
on the observed number of air shower particles.  The systematic
pointing error is also estimated to be smaller than $0\fdg011$.  These
are verified by the Moon's shadow observation \cite{Ame09}.

Using the Tibet-III and previous arrays, we have successfully observed
TeV $\gamma$-ray sources, such as the Crab Nebula \cite{Ame99,Ame09},
Mrk~501 \cite{Ame00}, and Mrk~421 \cite{Ame03}. We have also
successfully drawn a precise two-dimensional map of the large-scale
cosmic-ray anisotropy in the northern sky \cite{Ame06}, where we first
pointed out new small-area enhancements in the Cygnus arm direction at
multi-TeV energies.  One of the enhancements is coincident with MGRO~J2019+37,
which was established recently by the Milagro experiment as a TeV $\gamma$-ray
source \cite{Abd07}.  It is worth noting that the Tibet AS$\gamma$
experiment has reported several times \cite{Zha03,Zha05a,Zha05b} on
the marginal excesses from the direction closing to MGRO~J1908+06/HESS
J1908+063 before the final discovery made by the Milagro experiment.

\section{Air Shower Data Analysis}
We analyze the air shower dataset collected by the Tibet-III array during 1915.5 live days from 1999
November through 2008 December.  To
extract an excess of TeV $\gamma$-ray air shower events coming from
the direction of a target source in this analysis, we adopt almost the
same event selections and the background estimation method published
in our previous work \cite{Ame03,Ame09}.  We use air shower events
with $\sum\rho_{\rm FT}>10^{1.25}$ as the primary energy reference,
where the size $\sum\rho_{\rm FT}$ is defined as the sum of the number
of particles per m$^2$ for each FT detector.  The modal $\gamma$-ray
energy, assuming the Crab's orbit and integral spectral index $-$1.6, is
calculated to be approximately 3~TeV by the Monte Carlo simulation.
The modal $\gamma$-ray energy depends on the declination and is
estimated to be $\sim$3~TeV for a declination band from 20$\degr$ to
40$\degr$ and $\sim$6~TeV for declinations at 0$\degr$ and 60$\degr$,
respectively.  The search window radius centered at the target source
is expressed by $R(\sum\rho_{\rm FT}) = 6.9/\sqrt{ \sum\rho_{\rm FT}
}$ degrees, which is shown to maximize the signal-to-noise (S/N) ratio
by Monte Carlo study assuming a point-like $\gamma$-ray source
\cite{Ame03}.  Therefore, an excess might be underestimated if the
target source actually extends beyond our angular resolution size.

The target sources in the {\it Fermi} bright source list are chosen as
confirmed or potential Galactic sources in a similar way to that employed by
the Milagro observation \cite{Abd09b}.  Out of the 205 most
significant sources in the {\it Fermi} bright source list, 83 are not
identified as extragalactic sources \cite{Abd09a}.  Among these 83, we
select 27 sources in the declination band between 0$\degr$ and
60$\degr$, corresponding to our sensitive field of view for TeV
$\gamma$-ray sources.

\section{Results and Discussion}

The Tibet-III array observation of the selected 27 {\it Fermi} bright
Galactic sources is summarized in Table~\ref{tbl-1}, where 13 of the
selected sources are classified as pulsars (PSR), 5 are supernova
remnants (SNR), and 9 remain unidentified but are potential Galactic
sources, as they are mostly concentrated in the Galactic plane
($|b|<\sim20\degr$) \cite{Abd09a}.  As a result of this excess search
for these sources, we find no statistically significant evidence for
TeV $\gamma$-rays from other individual sources except for the Crab,
which is recognized as the brightest standard TeV source.

\begin{figure}
\epsscale{1.00}
\plotone{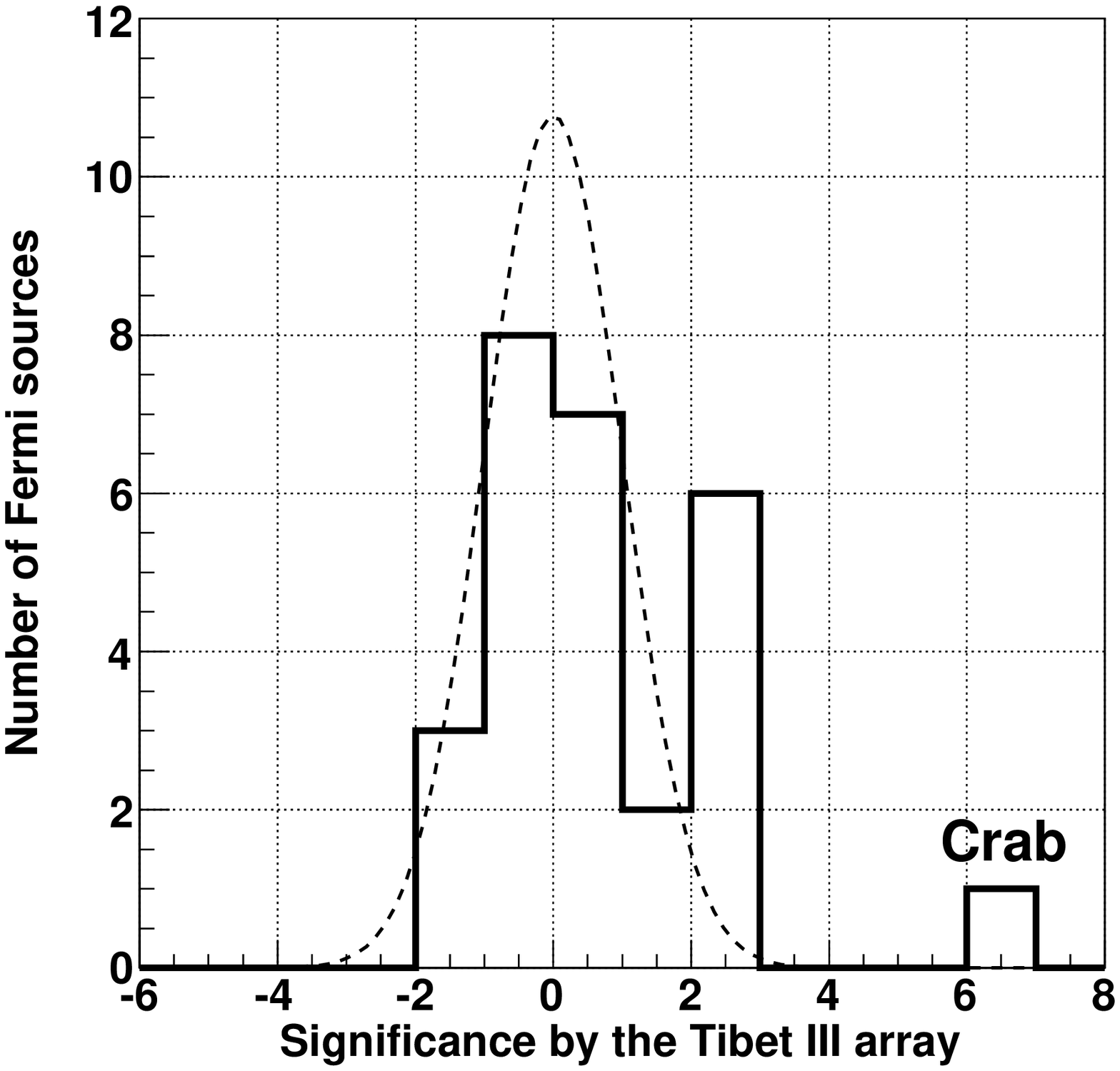}
\caption{ Histograms show significance distribution of the {\it Fermi}
  bright sources observed by the Tibet-III array.  The dashed curve
  is the expected normal Gaussian distribution.}
\label{fig1}
\end{figure}

Subsequently, the distribution of the observed significance is
examined for statistical consistency with the normal Gaussian.
Figure~\ref{fig1} shows the significance distribution of the 27
sources observed by the Tibet-III array.  One can see that the
$\gamma$-rays from the Crab are detected at a sufficiently high
significance of 6.9$\sigma$.  It should be emphasized that we observe
7 sources including the Crab at a significance of 2$\sigma$ or more in
this distribution, against an expected 0.61 sources (upper probability of
2$\sigma$ multiplied by 27 sources) from the normal Gaussian.  The
chance probability from Poisson statistics would be estimated as
3.8$\times$10$^{-6}$.  With the Crab excluded, the chance probability
would be estimated as 3.6$\times$10$^{-5}$.  This low chance
probability clearly shows that the {\it Fermi} bright Galactic sources
have statistically significant correlations with the TeV $\gamma$-ray
excesses.  In order to check the possible bias of the data sample,
spatially independent dummy sources are selected from the northern sky
except for the Crab, Mrk~421, and two famous large-scale cosmic-ray
anisotropy regions known as the Loss-Cone and the Tail-In regions
\cite{Ame06}.  The Loss-Cone and the Tail-In regions are separated
from the selected 27 {\it Fermi} sources.  As a result, the
significance distribution of $\sim$2000 dummy sources is consistent
with the normal Gaussian with a mean value of $m=-0.010\pm0.025$ and a
standard deviation of $\sigma=1.027\pm0.019$.
If the significance distribution observed by the Tibet-III array has a
density gradient toward the Galactic plane, the expected number of
sources at 2$\sigma$ or more may be enhanced in chance
association.  To check this, we perform a simple Monte Carlo
simulation in a similar way to that implemented by Romero et al.\ (1999). We
generate 2000 dummy-source lists of the 27 {\it Fermi} bright Galactic
sources, where the Galactic latitude distributions retain the form of
the actual histograms of the {\it Fermi} sources with the Galactic
longitudes randomly set to new distributions within our field of view.
We count the number of $\ge$2$\sigma$ sources in the Tibet-III data
according to each dummy-source list, and calculate that the average expected
number of sources at 2$\sigma$ or more is 
0.73$\pm$0.02 in the chance association.  In this case, the chance
probability associated with 7 sources or more goes up slightly, to
1.2$\times$10$^{-5}$, while it becomes 9.7$\times$10$^{-5}$ with the
Crab excluded.

The Milagro observation found 14 out of 34 {\it Fermi} sources at a
significance of 3$\sigma$ or more, and its sensitivity is
approximately two or three times better than that of the Tibet-III
array. Hence, our threshold significance 2$\sigma$, which corresponds
to $\sim$30\% of the Crab flux assuming a point-like source, should be
a reasonable value. We note that the flux of 0FGL~J2020.8+3649
seems to be quite low compared with the Milagro's, since the
flux measurement of 0FGL~J2020.8+3649 with the Milagro experiment is
(67$\pm$7)\% of the Crab flux above 35~TeV, while our flux is
(30$\pm$14)\% of the Crab flux above 3~TeV.  The statistical
difference between them is calculated to be 2.3$\sigma$.  This difference may be
explained by either statistical fluctuations, a harder energy
spectrum than that of the Crab, or an extended source instead of the
assumed point-like source in this analysis.

We also find that all 7 sources at 2$\sigma$ or more are associated with pulsars, and 6 of
them are coincident with sources detected by the Milagro experiment at
a significance of 3$\sigma$ or more at the representative energy of 35~TeV.
The remaining source still has a positive significance of 1.4$\sigma$
measured by the Milagro experiment.  Furthermore, the latest
{\it Fermi} LAT observations detected 16 $\gamma$-ray pulsars in the
blind frequency searches \cite{Abd09c}, among which only
LAT~PSR~J2238+59 is a new pulsar not included in the {\it Fermi}
bright source list. We also find a 2.5$\sigma$ excess associated with
LAT~PSR~J2238+59, as listed in Table~\ref{tbl-1}.  The location of this pulsar is
observed at a significance of 4.7$\sigma$ by the Milagro experiment
\cite{Abd09b}.
In this connection, the first {\it Fermi} LAT catalog including 46
$\gamma$-ray pulsars has been published using the first 6 months of
data \cite{Abd09d}
\footnote[28]{Note added in press: in this catalog, 18 pulsars are
  located in the Tibet-III field of view. Among them, 14 pulsars are
  listed in Table~\ref{tbl-1} including LAT~PSR~J2238+59.  We find no
  significant excess at 2$\sigma$ or more from the other 4 pulsar
  locations not listed in Table~\ref{tbl-1}.  Using the first {\it Fermi} 
  LAT pulsar catalog, we find that 8 out of 18 pulsar locations have
  a significance of 2$\sigma$ or more. In this case, the chance
  probability falls to 1.4$\times$10$^{-8}$, while it becomes
  1.8$\times$10$^{-7}$ with the Crab excluded.}.
We remark that no pulsed emission has ever been detected from
$\gamma$-ray pulsars above a few tens of GeV \cite{Ali08}.

\begin{figure*}
\epsscale{0.65}
\plotone{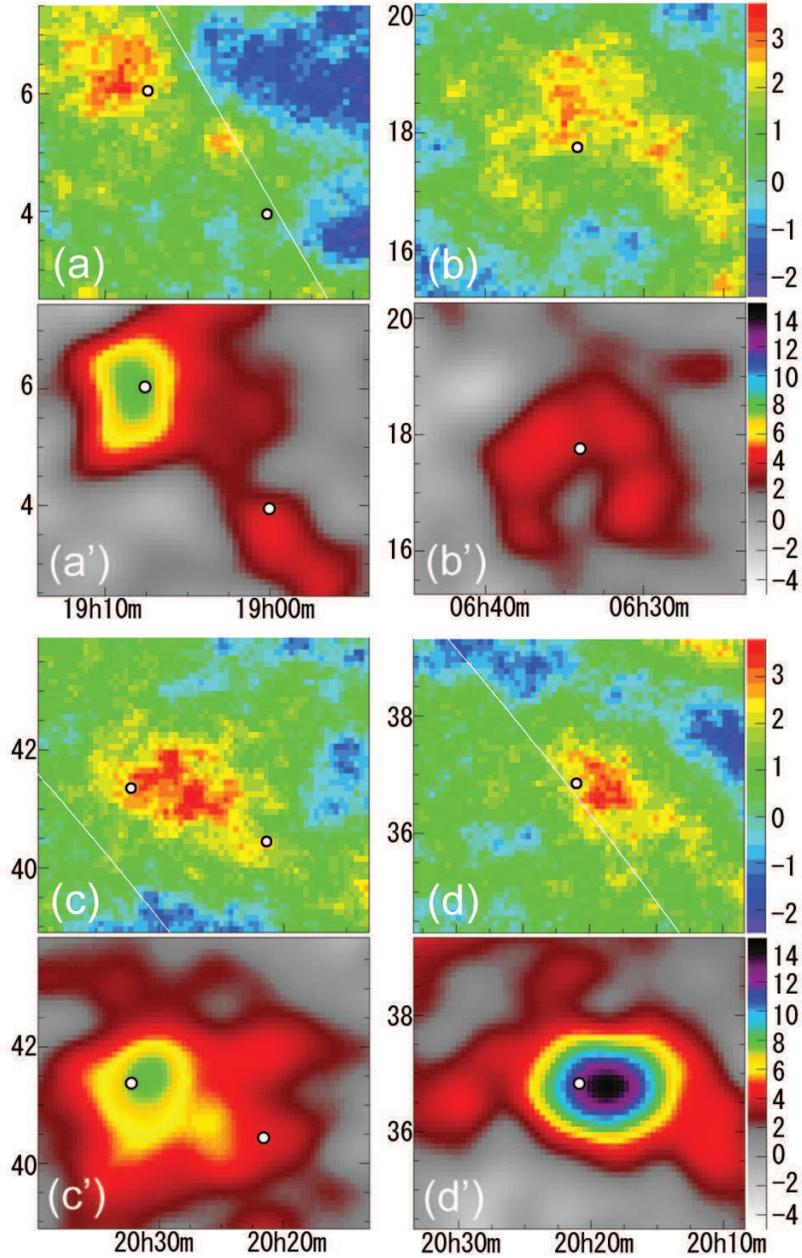}
\caption{
Comparisons of significance maps around the {\it Fermi} sources
between the Tibet-III array (a)--(d) and the Milagro experiment
(a')--(d') taken from Abdo et al. (2009c).  Selected are {\it Fermi}
sources with $\ge$2$\sigma$ significance by the Tibet-III array and $\ge$3$\sigma$
by the Milagro experiment except for the Crab.  White points in each image show
the {\it Fermi} source positions:
(a)(a') J1907.5+0602/J1900.0+0356;
(b)(b') J0634.0+1745 (Geminga);
(c)(c') J2021.5+4026/J2032.2+4122;
(d)(d') J2020.8+3649.
The horizontal axis, vertical axis, and color contours indicate the right
ascension, declination, and significance, respectively.}
\label{fig2}
\end{figure*}

Around the {\it Fermi} Galactic bright sources with $\ge$2$\sigma$
significance by the Tibet-III data and $\ge$3$\sigma$ by the Milagro
data except for the Crab, we compare significance maps between the
Tibet-III array (a)--(d) and the Milagro experiment (a')--(d')
taken from Abdo et al.\ (2009b) in Figure~\ref{fig2}.  Each image has
a $5\degr\times5\degr$ region including one or two {\it Fermi}
sources.  It is remarkable that the Tibet-III array obtains images consistent
with those observed in the Milagro experiment.  Besides, the
maximum significance positions obtained by both the Tibet-III array
and the Milagro experiment might be shifted from the pulsar positions.
In fact, recent imaging air Cherenkov telescopes also discovered many
candidates for TeV pulsar wind nebulae (PWNe), which are displaced
within a few tenths of degree from the pulsars in the southern sky
\cite{Aha06,Aha07}.  Thus, these observations would imply that the
excesses are possible candidates for TeV PWNe.  The correlation
between TeV and GeV $\gamma$-rays is being realized by the wide sky
survey instruments, such as the Tibet-III array and the Milagro
experiment, in the early {\it Fermi} era.

\acknowledgments 
The collaborative experiment of the Tibet Air Shower Arrays has been
performed under the auspices of the Ministry of Science and Technology
of China and the Ministry of Foreign Affairs of Japan. This work was
supported in part by a Grant-in-Aid for Scientific Research on
Priority Areas from the Ministry of Education, Culture, Sports,
Science and Technology, by Grants-in-Aid for Science Research from the
Japan Society for the Promotion of Science (JSPS) in Japan, and by the
Grants from the National Natural Science Foundation of China, the
Chinese Academy of Sciences, and the Ministry of Education of China.
K.~Kawata is supported by Grant-in-Aid for JSPS Fellows 21$\cdot$9437.

\begin{deluxetable}{ccccccc}
\tabletypesize{\small}
\tablecaption{Summary of the Tibet-III array observations of the {\it Fermi} sources. \label{tbl-1}}
\tablewidth{0pt}
\tablehead{
\colhead{{\it Fermi} LAT} & & & & \colhead{Tibet-III} & \colhead{Milagro\tablenotemark{a}} & \colhead{Source}\\
\colhead{Source} & \colhead{Class} &  \colhead{R.A.}  &  \colhead{Dec.}  & \colhead{Signi.} & \colhead{Signi.} & \colhead{Associations} \\
\colhead{(0FGL)} & & \colhead{(deg)} & \colhead{(deg)} & \colhead{($\sigma$)} & \colhead{($\sigma$)}  &            
}
\startdata
J0030.3+0450 & PSR & 7.600  & 4.848   & 1.7  & $-$1.7 &   \\
J0357.5+3205 & PSR\tablenotemark{b} & 59.388 & 32.084  & $-$1.7 & $-$0.1 &   \\
J0534.6+2201 & PSR & 83.653 & 22.022  & 6.9  & 17.2 & Crab  \\
J0617.4+2234 & SNR & 94.356 & 22.568  & 0.2  & 3.0  & IC 443  \\
J0631.8+1034 & PSR & 97.955 & 10.570  & 0.3 & 3.7  &   \\
J0633.5+0634 & PSR\tablenotemark{b} & 98.387 & 6.578   & 2.4  & 1.4  &   \\
J0634.0+1745 & PSR & 98.503 & 17.760  & 2.2  & 3.5  & Geminga  \\
J0643.2+0858 &     & 100.823 & 8.983  & $-$1.2 & 0.3  &   \\
J1830.3+0617 &     & 277.583 & 6.287  & $-$0.2 & 0.2  &   \\
J1836.2+5924 & PSR\tablenotemark{b} & 279.056 & 59.406 & $-$0.3 & $-$0.9 &   \\
J1855.9+0126 & SNR & 283.985 & 1.435  & 0.7  & 2.2  & W44  \\
J1900.0+0356 &     & 285.009 & 3.946  & 1.0  & 3.6  &   \\
J1907.5+0602 & PSR\tablenotemark{b} & 286.894 & 6.034  & 2.4  & 7.4  & MGRO~J1908+06 \\
& & & & & & HESS~J1908+063  \\
J1911.0+0905 & SNR & 287.761 & 9.087  & 1.7  & 1.5  & G43.3$-$0.2  \\
J1923.0+1411 & SNR & 290.768 & 14.191 & $-$0.3 & 3.4  & W51 \\
& & & & & & HESS~J1923+141  \\
J1953.2+3249 & PSR & 298.325 & 32.818 & $-$0.0 & 0.0  &   \\
J1954.4+2838 & SNR & 298.614 & 28.649 & 0.6  & 4.3  & G65.1+0.6  \\
J1958.1+2848 & PSR\tablenotemark{b} & 299.531 & 28.803 & 0.1  & 4.0  &   \\
J2001.0+4352 &     & 300.272 & 43.871 & $-$0.5 & $-$0.9 &   \\
J2020.8+3649 & PSR & 305.223 & 36.830 & 2.2  & 12.4 & MGRO~J2019+37  \\
J2021.5+4026 & PSR\tablenotemark{b} & 305.398 & 40.439 & 2.2  & 4.2  &   \\
J2027.5+3334 &     & 306.882 & 33.574 & $-$0.3 & $-$0.2 &   \\
J2032.2+4122 & PSR\tablenotemark{b} & 308.058 & 41.376 & 2.4  & 7.6  & TeV~J2032+4130 \\
& & & & & & MGRO~J2031+41 \\
J2055.5+2540 &     & 313.895 & 25.673 & $-$0.0  & $-$0.0 &   \\
J2110.8+4608 &     & 317.702 & 46.137 & 0.3  & 1.1  &   \\
J2214.8+3002 &     & 333.705 & 30.049 & $-$1.0 & 0.6  &   \\
J2302.9+4443 &     & 345.746 & 44.723 & $-$0.0 & $-$0.6 &   \\
\tableline\tableline
LAT PSR J2238+59\tablenotemark{c} & PSR\tablenotemark{b}  & 339.561 & 59.080 & 2.5 & 4.7 &  \\
\enddata
\tablenotetext{a}{Significance of the Milagro observations. Taken from \cite{Abd09b}.}
\tablenotetext{b}{These pulsars are newly discovered by the {\it Fermi} LAT 
observations \cite{Abd09a,Abd09c}.}
\tablenotetext{c}{This pulsar is not in the {\it Fermi} bright source list, 
but it is detected by the latest {\it Fermi} LAT observation \cite{Abd09c}.}
\end{deluxetable}






\end{document}